\documentclass[10pt,notitlepage,pra,twocolumn]{revtex4}
\usepackage{amsfonts}
\usepackage{amsmath}
\usepackage{amssymb}
\usepackage{graphicx}
\usepackage[toc,page,header]{appendix}

\providecommand{\U}[1]{\protect\rule{.1in}{.1in}}

\begin{document}

\title{Dissipation-sensitive multi-photon excitations of strongly interacting Rydberg atoms}
\author{Jing Qian and Weiping Zhang}
\affiliation{Quantum Institute for Light and Atoms, Department of Physics, East China
Normal University, Shanghai 200062, People's Republic of China}

\begin{abstract}
We theoretically investigate the effect of dissipation on multi-photon excitation of Rydberg atoms. The steady states and the dynamics are compared  via two types of four-level excitation schemes with different dissipative paths of spontaneous emission. We find that in the case of strong Rydberg-Rydberg interaction, the schemes will settle in several different non-equilibrium steady states. The interesting aspect is that there exists the multi-stable steady states, which reveals the competition between interaction-induced nonlinearity and dissipation 
caused by spontaneous emission. A numerical simulation on the Rydberg population dynamics in the bistable region exhibits different features existing in the two schemes even with the same initial conditions, which accounts for the influence of the dissipation on the dynamics.
\end{abstract}

\maketitle
\preprint{}

\section{Introduction}

Due to the large electric dipole moments, there are strong and long-range interactions between Rydberg atoms and these interactions can be controlled and enhanced by external electromagnetic fields. 
Besides, the huge polarizability of Rydberg states gives rise to giant Kerr coefficients \cite{Mohapatra08}, allowing nonlinear optical effects for only a few photons \cite{Peyronel12}.
That is why the atoms excited to the high-lying Rydberg states are of great interest in the recent researches of quantum many-body physics \cite{WeimerS}, quantum information processing \cite{Saffman10} and quantum nonlinear optics \cite{Peyronel12}.

The high-lying Rydberg states can be a bit more easily excited from atomic ground states by laser light via absorption of more than one photon \cite{Carr12} , while the spontaneous emission of the intermediary states will introduce the dissipative mechanism into the Rydberg excitation schemes in spite of the long life of the Rydberg states \cite{Olmos14}. In general, these environment-induced dissipations inevitably lead to decoherence and noise in the quantum system and thus are undesirable in the above quantum physics researches. 

In recent years, however, 
the dissipation has been altered its role into a useful resource and tool for a lot of
quantum applications, such as dissipative state engineering in trapped ions \cite{Barreiro11,Genway14} and cold atoms \cite{Diehl08}, dissipative quantum
computation \cite{Verstraete09}, dissipative quantum optics \cite{Gorshkov13}, bound state formation in molecules \cite{Lemeshko13} and entangled steady state production \cite{Lin13,Rao13,Carr13L}. 
The key for these applications is the occurrence of ``non-equilibrium stationary state'' which can be achieved when the driving and dissipative processes arrive at a dynamical equilibrium \cite{Diehl10,Dalla10,Lesanovsky13,Petrosyan13}. 
By appropriately arranging the system-environment couplings, these steady states can deviate far from the classical thermal equilibrium states and keep unique quantum features resisting decoherences \cite{Lesanovsky14}.
Moreover, combining with the strong dipole-dipole interaction and blockade mechanics of the Rydberg atoms, the dissipation could give rise to quite a few exotic phases of ultracold many-body atomic system, such as uniform phase, antiferromagetic phase, oscillatory phase and even the bistability between these phases \cite{Tony11,Qian12,Hu13}.

In this work we investigate the effect of dissipation on the steady states and dynamics of multi-photon Rydberg excitation schemes.
To focus on the dissipation induced by the spontaneous emission of the intermediary states, we choose two typical four-level schemes: one level structure is $N$-type and the other is cascade. By appropriately arranging the laser fields, these two schemes have exactly the same coherent dynamics. However, if including the effect of dissipation their realistic dynamics are of great difference, since a dissipative channel due to sponaneous decay running in the opposite direction in the two schemes.
We compare their non-equilibrium steady states by solving the
stationary master equations under the mean-field treatment, and find that the
number and distribution of their steady state solutions are entirely different
when the Rydberg nonlinear interaction is strong. 
The bistable steady states are present in the the $N$-type scheme while the Autler-Townes Splitting (ATS) occurs in the cascade scheme resulting in the tri-stable steady states. 
As a consequence, for a strong nonlinear system such as Rydberg atom systems,
dissipative force should be considered very carefully since the final steady states of the system will be very sensitive to it. 
Furthermore, we study the
dynamical evolution of the two schemes in the bistable region. Due to the difference in the dissipative channel, we observe two schemes may evolute into different
branches under the same initial conditions, which makes sense for dissipative preparation of quantum state
in strong nonlinear interacting systems.

\section{Model and Master Equation}

\begin{figure}
\includegraphics[scale=0.45]{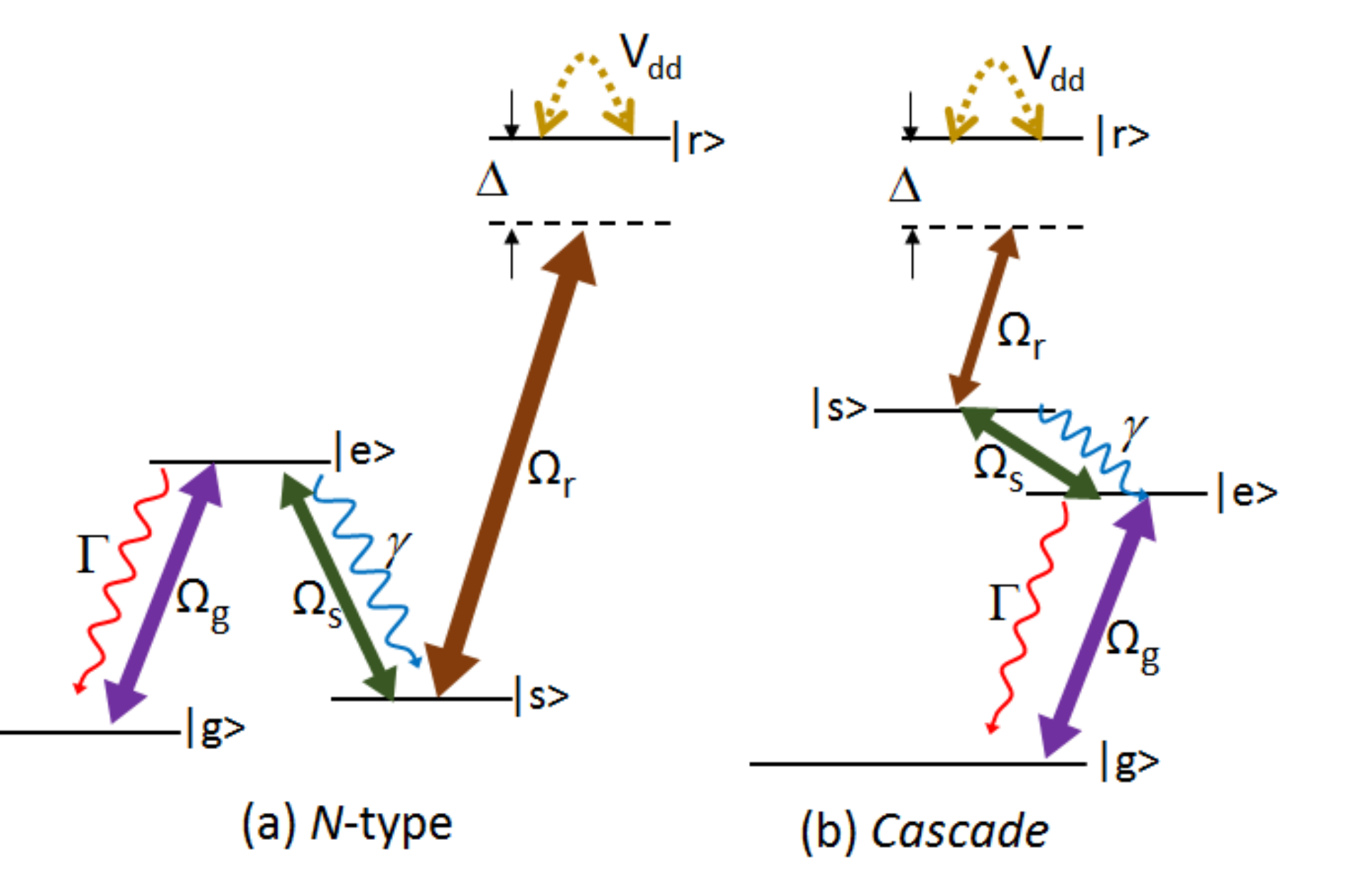}
\caption{(Color online) $N$-type and
cascade atomic multi-level systems. States $\left\vert g\right\rangle $, $%
\left\vert e\right\rangle $, $\left\vert s\right\rangle $ and $\left\vert
r\right\rangle $ respectively show the ground state, two intermediate states
and Rydberg state, which are successively coupled by laser Rabi frequencies $%
\Omega _{g}$, $\Omega _{s}$ and $\Omega _{r}$. The three-photon laser
detuning $\Delta $ can be controlled in experiment. Other parameters are
described in the text.}
\label{model_scheme}
\end{figure}

In what follows, we will discuss two different multi-step cw-excitation schemes for the Rydberg state, as typically used in experiments. One is an $N$-type scheme with a Raman transition before a Rydberg-excited transition, as shown in Fig. \ref{model_scheme}(a). The atom is first excited from its ground state $\vert g\rangle$ to an intermediate state $\vert e\rangle$ with a transition strength given by the Rabi frequency $\Omega_g$, and then transits to a metastable state $\vert s\rangle$ with Rabi frequency $\Omega_s$. Finally, another laser drives the transition between the metastable level and the desired Rydberg state $\vert r\rangle$ with Rabi frequency $\Omega_r$. 
The photon for the Raman transition is typically provided by the MOT lasers, which are tuned on resonance with the levels during the time of Rydberg excitation. So the spontaneous emissions from the intermediate state $\vert e\rangle$ to state $\vert g\rangle$ with rate $\Gamma$ and to state $\vert s\rangle$ with rate $\gamma$, respectively, have to be taken into account. 
On the other hand, we
allow for a detuning $\Delta$ from resonant Rydberg excitation and the lifetime of the Rydberg state is much longer (typically as large as tens of $\mu $s with the principle quantum number $n\sim 50$) so that its decay can be safely neglected in our discussion. 
The second scheme shown in Fig. \ref{model_scheme}(b) has a cascade level structure, where $\vert s\rangle$ represents the higher intermediate state, so that $\gamma$ is the decay rate from state $\vert s\rangle$ to state $\vert e\rangle$, opposite to the case in the first scheme. The physical meanings of the other quantities remain unchanged. Besides, we note that in this scheme decay from state $\left\vert s\right\rangle $ to state $\left\vert g\right\rangle $ is physically forbidden according to the transition selection rule. 
Except the atom-light interaction, we also consider the strong interaction between Rydberg atoms (Here we only consider a homogeneous van-der-Waals interaction), which under the mean field approximation will cause a population-dependent energy shift to the Rydberg state. 

Mean field theory is a classical approximation to the quantum model in which quantum correlations between atoms are ignored. Note that it works well only with large atom number \cite{Carr13}. For systems of several Rydberg atoms, the quantum effect of the strong Rydberg blockade can no longer be effectively described by the mean field approximation \cite{Urban09,Gaetan09,Barredo14,Hakin14}. In our consideration, we assume that the number of atom is large enough so that the mean field approach can be safely used.

In frames rotating at appropriate frequencies respectively, these two different Rydberg-excitation schemes can be described by a same mean-filed Hamiltonian ($\hbar =1$),
\begin{equation}
\mathcal{H}=\left( \Delta +V_{dd}\rho _{rr}\right) \sigma _{rr}+(\Omega
_{g}\sigma _{ge}+\Omega _{s}\sigma _{es}+\Omega _{r}\sigma _{sr}+\text{H.c.}),
\label{Ham}
\end{equation}
with $\sigma _{ij}=\left\vert i\right\rangle \left\langle j\right\vert $
being the corresponding transition operator between two internal atomic
states, $V_{dd}$ the Rydberg interaction strength, and $\rho _{rr}$ the Rydberg state population.
The only difference is the expression of the three-photon detuning, given by $\Delta =\omega _{r}-\omega _{\Omega _{g}}\mp
\omega _{\Omega _{s}}-\omega _{\Omega _{r}}$ with frequencies $\omega _{r}$%
, $\omega _{\Omega _{g}}$, $\omega _{\Omega _{s}}$, $\omega _{\Omega _{r}}$
denoting the energy of state $\left\vert r\right\rangle $, the carrier
frequencies of the fields $\Omega _{g}$, $\Omega _{s}$, $\Omega _{r}$,
respectively. The upper sign is for the $N$-type scheme and the lower one for the cascade scheme.
That means the coherent dynamics of these two schemes can be same even they have different structures of energy levels, provided the three-photon detunings are equal. 
However, we will show below in a realistic experiment where the dissipation caused by the spontaneous decay should be considered, these schemes will display distinct non-equilibrium steady states and dynamics, which is attributed to the opposite direction of decay route between level $\left\vert s\right\rangle $ and level $\left\vert e\right\rangle $ in two schemes. 

In order to investigate the influences of the spontaneous emissions from the intermediate excited states,
we use the Lindblad master equation \cite{Qian13}
\begin{equation}
\partial _{t}\hat{\rho}=-i\left[ \mathcal{H},\hat{\rho}\right] +\mathcal{L}\left[ \hat{\rho}\right],  \label{master_eq}
\end{equation}%
where $\Hat{\rho}$ is the single-atom density matrix operator which can well capture the evolution of the Rydberg atoms according to the mean field approximation, and $\mathcal{L}\left[ 
\hat{\rho}\right] $ is the Lindblad operator which is introduced
phenomenologically to depict the atomic spontaneous emissions due to optical
excitations \cite{Breuer02}:%
\begin{equation}
\mathcal{L}\left[ \Hat{\rho}\right] =\frac{\Gamma }{2}\left( 2\sigma
_{ge}\Hat{\rho}\sigma _{ge}^{\dagger }-\left\{ \sigma _{ee},\Hat{\rho}%
\right\} \right) +\frac{\gamma }{2}\left( 2d\Hat{\rho}d^{\dagger }-\left\{
d^{\dagger }d,\Hat{\rho}\right\} \right)
\end{equation}%
where ${\{A , B\}}=A B +B A$, and $d=\sigma _{se}$ for the $N$-type system and $\sigma _{es}$ for the
cascade system. From that we can obtain two different groups of differential equations of the density matrix elements $\rho_{ij}$. 
For the $N$-type system we have, 
\begin{eqnarray}
\Dot{\rho}_{ee} & = & 2\Omega_{g}\text{Im}( \rho_{ge}) - 2\Omega_{s}\text{Im}\left( \rho _{es}\right) -\left( \gamma +1\right) \rho_{ee},  \notag \\
\Dot{\rho}_{ss} & = & 2\Omega _{s}\text{Im}\left( \rho _{es}\right) -2\Omega_{r}\text{Im}\left( \rho _{sr}\right) +\gamma \rho _{ee},  \notag \\
\Dot{\rho}_{rr} & = & 2\Omega _{r}\text{Im}\left( \rho _{sr}\right) ,\notag \\
\Dot{\rho}_{ge} & = & -i\Omega _{g}\left( \rho _{ee}-\rho _{gg}\right) +i\Omega_{s}\rho _{gs}-\frac{\gamma +1}{2}\rho _{ge} , \notag \\
\Dot{\rho}_{gs} & = & -i\Omega _{g}\rho _{es}+i\Omega _{s}\rho _{ge}+i\Omega_{r}\rho _{gr} ,\notag \\
\Dot{\rho}_{gr} & = & i\Delta _{eff}\rho _{gr}-i\Omega _{g}\rho _{er}+i\Omega_{r}\rho _{gs},  \notag \\
\Dot{\rho}_{es} &=& -i\Omega _{g}\rho _{gs}-i\Omega _{s}\left( \rho_{ss}-\rho _{ee}\right) +i\Omega _{r}\rho _{er}-\frac{\gamma +1}{2}\rho _{es},
\notag \\
\Dot{\rho}_{er} &=& (i\Delta _{eff}-\frac{\gamma +1}{2})\rho _{er}-i\Omega_{g}\rho _{gr}-i\Omega _{s}\rho _{sr}+i\Omega _{r}\rho _{es},  \notag \\
\Dot{\rho}_{sr} &=& i\Delta _{eff}\rho _{sr}-i\Omega _{s}\rho _{er}-i\Omega_{r}\left( \rho _{rr}-\rho _{ss}\right) , \label{Ntypef}
\end{eqnarray}
while for the cascade system, 
\begin{eqnarray}
\Dot{\rho}_{ee} &=&2\Omega _{g}\text{Im}\left( \rho _{ge}\right) -2\Omega
_{s}\text{Im}\left( \rho _{es}\right) -\rho _{ee}+\gamma \rho _{ss} , \notag
\\
\Dot{\rho}_{ss} &=&2\Omega _{s}\text{Im}\left( \rho _{es}\right) -2\Omega
_{r}\text{Im}\left( \rho _{sr}\right) -\gamma \rho _{ss} , \notag \\
\Dot{\rho}_{rr} &=&2\Omega _{r}\text{Im}\left( \rho _{sr}\right)  \notag \\
\Dot{\rho}_{ge} &=&-i\Omega _{g}\left( \rho _{ee}-\rho _{gg}\right) +i\Omega
_{s}\rho _{gs}-\frac{1}{2}\rho _{ge} , \notag \\
\Dot{\rho}_{gs} &=&-i\Omega _{g}\rho _{es}+i\Omega _{s}\rho _{ge}+i\Omega
_{r}\rho _{gr}-\frac{\gamma }{2}\rho _{gs} , \notag \\
\Dot{\rho}_{gr} &=&i\Delta _{eff}\rho _{gr}-i\Omega _{g}\rho _{er}+i\Omega
_{r}\rho _{gs} , \notag \\
\Dot{\rho}_{es} &=&-i\Omega _{g}\rho _{gs}-i\Omega _{s}\left( \rho
_{ss}-\rho _{ee}\right) +i\Omega _{r}\rho _{er}-\frac{\gamma +1}{2}\rho _{es},
\notag \\
\Dot{\rho}_{er} &=&(i\Delta _{eff}-\frac{1}{2})\rho _{er}-i\Omega _{g}\rho
_{gr}-i\Omega _{s}\rho _{sr}+i\Omega _{r}\rho _{es} , \notag \\
\Dot{\rho}_{sr} &=&(i\Delta _{eff}-\frac{\gamma }{2})\rho _{sr}-i\Omega
_{s}\rho _{er}-i\Omega _{r}\left( \rho _{rr}-\rho _{ss}\right),
\label{cascadeTyf}
\end{eqnarray}
where we have assumed that the off-diagonal coherence elements $\rho_{ij}=\rho _{ji}^{\ast }(i\neq j)$ 
and the conservation of the total population that the diagonal elements $\rho _{gg}+\rho
_{ee}+\rho _{ss}+\rho _{rr}=1$. The effective detuning $\Delta _{eff}=\Delta +V_{dd}\rho _{rr}$, including the population-dependent frequency shift of the Rydberg state, which makes the differential equations nonlinear.
We can find these two groups of equations become identical when the decay $\gamma \rightarrow 0$, which means the dynamics of the two schemes are indistinguishable. Instead, a non-vanishing $\gamma$ will lead to entirely different excitation dynamics, especially in the regions with high nonlinearity induced by Rydberg-Rydberg interaction. To show these differences in the following we will keep the common decay $\Gamma$ constant and focus on the influence of the decay $\gamma$ which has opposite direction in the two schemes. Then it is also convenient to renormalize the time by $\Gamma^{-1}$ and all frequencies by $\Gamma$ in the equations.

\section{Steady states and phase diagram}
\begin{figure}
\includegraphics[width=3.3in,height=3.5in]{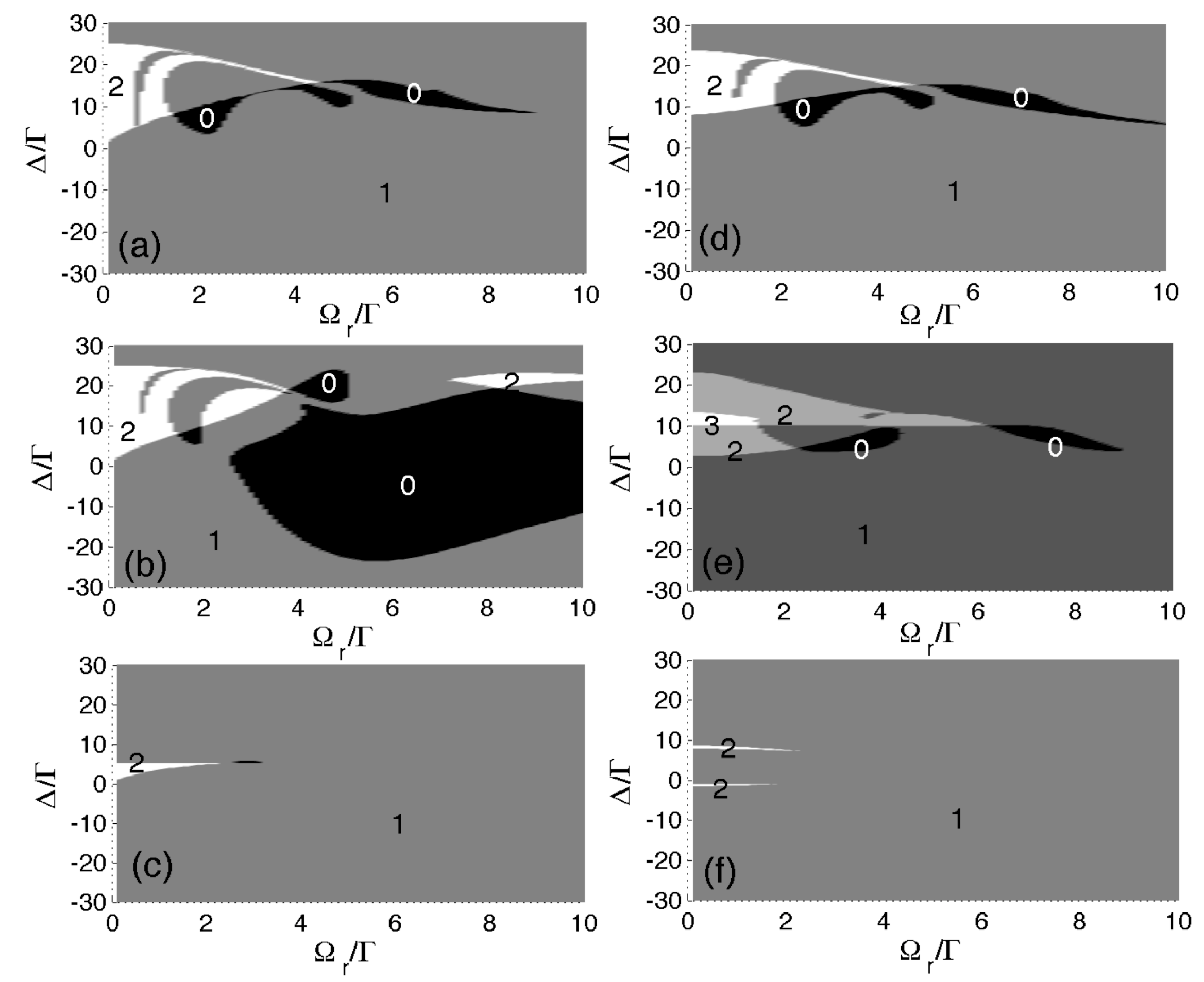}
\caption{Phase diagram in the $\left( \Omega_r ,\Delta \right) $ space,
indicating the influence of decay and nonlinearity on the phase of the Rydberg atoms, with the numbers of the stable solutions marked in different gray-level. The left column is for the $N$-type system and the right column is for the cascade one. 
From top to bottom, we fix the Rabi frequencies $\Omega _{g}=5.0$, $\Omega _{s}=2.0$, and change the strength of decay and Rydberg interaction: (a) and (d) $V_{dd}=-50$ and $\protect%
\gamma =0.01$; (b) and (e) $V_{dd}=-50$ and $\protect\gamma =1.0$; (c) and (f) $%
V_{dd}=-10$ and $\protect\gamma =1.0$. (All parameters are scaled by $\Gamma$.)}
\label{Phase_diagram}
\end{figure}

We first investigate the steady states of the schemes, where the competition between the oscillation caused by coherent optical driving and the dissipation due to the spontaneous emissions reaches a dynamical equilibrium so that the populations and the coherence for each level become stationary.  
The steady occupancy of the Rydberg state, $\rho_{rr}$, can then be readily obtained by setting all time differential terms in Eq. (\ref{Ntypef}) and (\ref{cascadeTyf}) equal zero and then solving the resulting algebraic equations.
For the $N$-type scheme, we obtain a cubic equation about $\rho_{rr}$ which may have one or three solutions corresponding to the uniform and the bistable phase of the Rydberg excitation scheme, respectively (We should check the stabilities of these solutions to ensure they are achievable in the dissipation environment). Here the bistable phase, as a typical nonlinear effect, is caused by the Rydberg-Rydberg interaction. It has been observed in recent experiment of a dilute Rydberg atomic ensemble \cite{Carr13}.
On the other side, the equation becomes quintic in the case of the cascade scheme, indicating the existence of multi-stable phase of the scheme, which hasn't been found in the Rydberg system. 
The analytical forms of these equations and their solutions are too cumbersome to present here, we instead show the phase diagram of the Rydberg atoms in the parameter space of $\Omega _{r}$ and $\Delta $, which is decided by the number of the stable solutions.  As shown in Fig. \ref{Phase_diagram}, the left column is for the $N$-type scheme and the right column for the cascade scheme,  with the regions labeled by numbers ``0''-``3" respectively corresponding to the oscillatory phase (none stable solution), the uniform phase (only one stable solution), bistable phase (two stable solutions) and tri-stable phase (three stable solutions). 

From the figure we can find the two schemes have similar phase diagrams at small decay strength (see the first row, $\gamma=0.01\Gamma$ and $V_{dd}=-50\Gamma$). The parameter space is dominated by the uniform phase, the bistable phase presents for small Rabi frequency $\Omega_r$, and the oscillatory phase only survives in a very narrow region where detuning $\Delta$ is small. For a large value of $\gamma$ (the second row, $\gamma=\Gamma$ and $V_{dd}=-50\Gamma$), the phase diagrams become distinct from each other. In the case of the $N$-type scheme the oscillatory phase occupies a large area with larger value of $\Omega_r$, that is because the non-lossy states, $\left\vert s\right\rangle $ and $\left\vert r\right\rangle $, sustain a Rabi-type population oscillation between them. Instead, in the case of the cascade scheme the decay of state $\left\vert s\right\rangle $ prevents such oscillation. Moreover, ATS of the cascade system causes a unique tri-stable phase (we will show it clearly below) which is absent in the former case \cite{Comparat10}.
We emphasize that the effect of dissipation-sensitive steady states is more remarkable under the condition of strong nonlinearity. When we decrease the Rydberg-Rydberg interaction strength, the phase diagrams of the two schemes trends to be analogous even for large value of decay (the third row, $\gamma=\Gamma$ and $V_{dd}=-10\Gamma$).  

\begin{figure}
\includegraphics[scale=0.36]{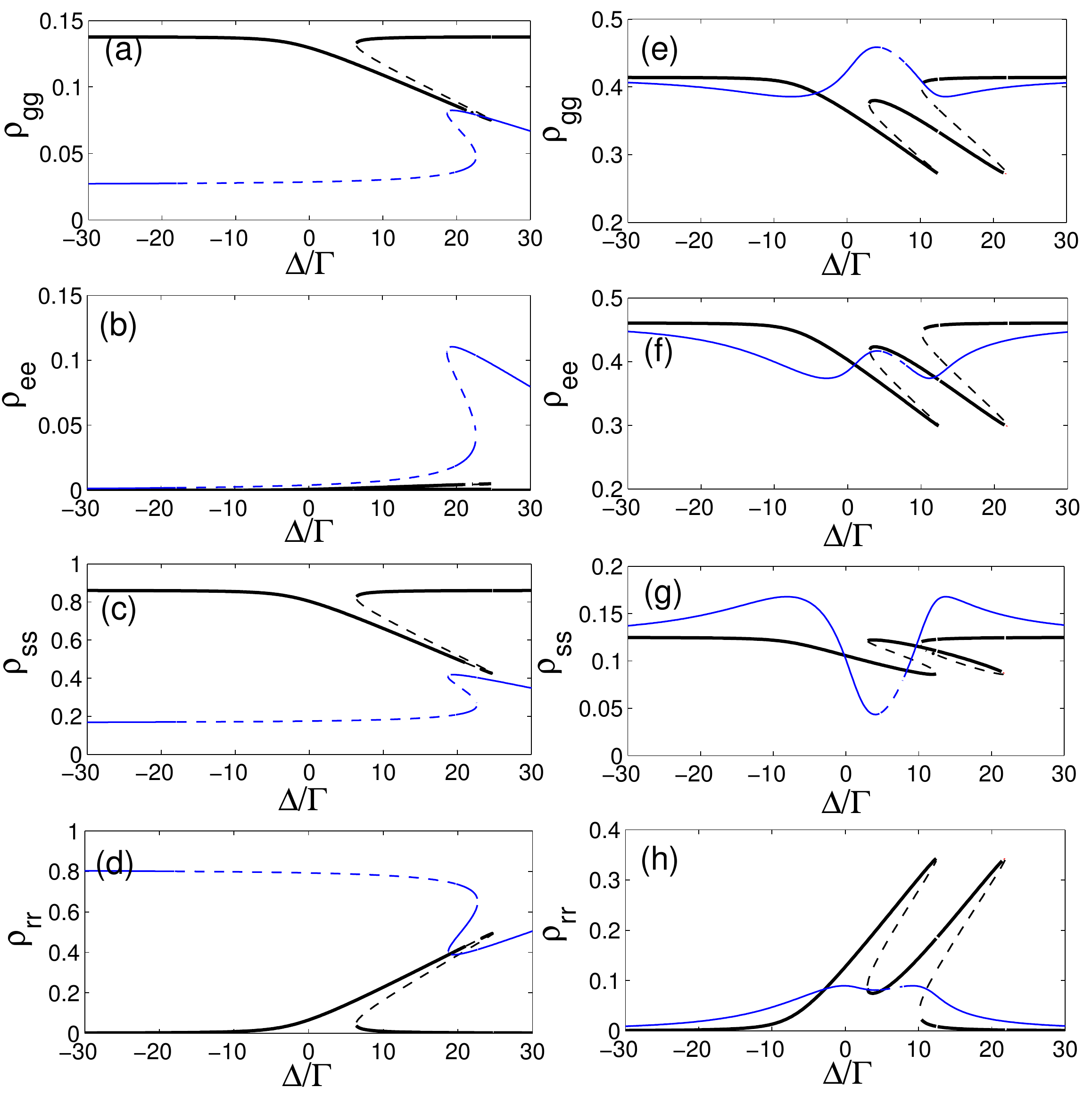}
\caption{(Color online) The steady state populations of state $|g\rangle$ (first row), state $| e\rangle$ (second row), state $| s \rangle$ (third row), and $| r\rangle$ (last row) are plotted as a function of three-photon detuning $\Delta $, with the weak Rydberg-state coupling case ($\Omega _{r}/\Gamma =1.0$) in thick black curves and the strong coupling case ($\Omega _{r}/\Gamma=8.0$) in thin blue curves. 
The dashed curves corresponds to the values in the unstable phases. 
The left and right columns are for the $N$-type system and the cascade system, respectively. 
Other parameters are same as Fig. \protect\ref{Phase_diagram}(b) and (e)}
\label{Rydberg_popu}
\end{figure}

Before focusing on the different quantum multi-stable phases of the two schemes in the parameter space of strong nonlinearity and large decay, we first investigate their steady states without nonlinearity, which will show the cause of these differences are the direction of the decay $\gamma$. By solving the stationary OBEs in the case of $V_{dd}=0 $ and other parameters same as Fig. \ref{Phase_diagram}(b) and (e), we observe that the Rydberg excitation 
spectrum in the cascade scheme shows a double-peak structure separated by $2\Omega_{g}$ which is same as the feature of the ATS presenting in the three-level Rydberg-excitation scheme \cite{Ate07, Amthor10}. 
In our $N$-type scheme, the same direction of decay $\gamma$ and $\Gamma$ makes the atoms populate mostly on state $\left\vert e\right\rangle$ and $\left\vert g\right\rangle$, resulting in the splitting of the Rydberg excitation spectrum. 
Moreover, different from in the three-level scheme where ATS is just an effect of transient stability \cite{Ate07A}, the existence of the intermediate state $\left\vert s\right\rangle$ here has stabilized this splitting and enabled it to appear in the final steady state of the scheme. In the case of the $N$-type scheme, the opposite directions of decays $\gamma$ and $\Gamma$, as well as small Rabi frequency $\Omega_s$, may accumulate the population on state $\left\vert s\right\rangle $ only, giving rise to a single peak instead of the ATS in the Rydberg excitation spectrum. These different peak structures still remain in the presence of Rydberg-Rydberg interactions. As we shown in Fig. \ref{Rydberg_popu}(a)-(h), the nonlinearity caused by the interaction distorts the original spectrum, resulting in the different multi-stable steady states of the cascade and the $N$-type schemes.

Again the right column of Fig. \ref{Rydberg_popu} is for the $N$-type scheme, where we display the exact values of the steady population of each states at small $\Omega _{r}$ case (thick black line) and large $\Omega _{r}$ case (thin blue line), respectively, corresponding to the two bistable regions marked in phase diagram Fig. \ref{Phase_diagram}(b). We can find different features of the bistable behaviors in these two cases. 
When $\Omega_{r}$ is small it exhibits intrinsic bistability in the region of $\Delta_{eff}=\Delta +V_{dd}\rho _{rr}\approx 0$ with its inclined direction of hysteresis window depending on the sign of Rydberg interaction $V_{dd}$. When $\Delta $
is negative or large positive (i.e. $\Delta _{eff}$ is off-resonance), the
ground states $\left\vert s\right\rangle $ is occupied with a dominant
number of atoms ($\rho _{ss}>0.8$). However, as increasing $\Omega _{r}$ to $%
8.0$ we find that the dominant population at $\left\vert \Delta \right\vert
\rightarrow \infty $ occupies the Rydberg state $\left\vert r\right\rangle $
with its value $\rho _{rr}\rightarrow \Omega _{g}^{2}\Omega _{r}^{4}/(\left(
\Omega _{g}^{2}+\Omega _{s}^{2}\right) ^{3}+\Omega _{g}^{2}\Omega _{r}^{4})$
due to a quite strong Rydberg excitation. Meanwhile, a small region of
bistable phase emerges near $\Delta /\Gamma =20$ where the Rydberg
interaction induced energy shift has been compensated by a large detuning
i.e. $\Delta _{eff}\approx 0$. Besides, when $\Omega_r$ is large, there is a large 
unstable region (corresponding to the oscillatory phase) denoted by the dashed lines, which is due to the strong
coupling between the non-lossy states $\left\vert s\right\rangle $ and $%
\left\vert r\right\rangle $.

Turning to the case of the cascade system (the right column of Fig. \ref{Rydberg_popu}),
a dramatic difference is the presence of the inclined double-peak configuration at small $\Omega_r$ case,
which gives rise to bistable and even tri-stable phases
when $\Delta $ and $\left\vert V_{dd}\right\vert \rho _{rr}$ are comparable.
In addition, state $\left\vert g\right\rangle $ and state $\left\vert e\right\rangle $ share the most atomic population ($\rho _{gg}+\rho
_{ee}>0.85$) and the case does not change much even for larger value of $\Omega _{r}$. However, the double-peak structure degenerates a lot so that there is no multi-stable phase in large $\Omega_r$ case. This is due to the comparable values of $\Omega_g$ and $\Omega_r$ prevent the occurrence of the ATS.

Summarizing, for the $N$-type scheme, decay $\gamma$ leads to a large population on state $\left\vert s\right\rangle $ which, combining with Rydberg state $\left\vert r\right\rangle $, makes the scheme work like a two-level system, so that the sustained Rabi-typed oscillation occurs with an intensive $\Omega _{r}$.
For small $\Omega _{r}$ the behavior of the scheme is more close to the
cooperative optical excitation \cite{Otterbach14}. For the cascade scheme, the existence of $%
\gamma $ leads to the ATS phenomenon, inducing multi-stable
steady state phases under the condition of strong nonlinear interaction.

\section{Bistability and Population dynamics}

The above phase diagrams reveal the properties of the steady states in the parameter space of detuning $\Delta$ and Rabi frequency $\Omega_r$, which are controlled by the external lasers. Specially, owing to the strong nonlinear interactions the bistable phases present, which suggests there are two different steady states, i.e., the low- and the high-Rydberg-occupied steady state, under the same laser parameters. In reality, the system would eventually evolute into one of them, which depends on the initial states and decay routes. In order to study this dependence, we perform a numerical simulation for the population dynamics by directly solving the OBEs (\ref{Ntypef}) and (\ref{cascadeTyf}).

\begin{figure}
\includegraphics[scale=0.45]{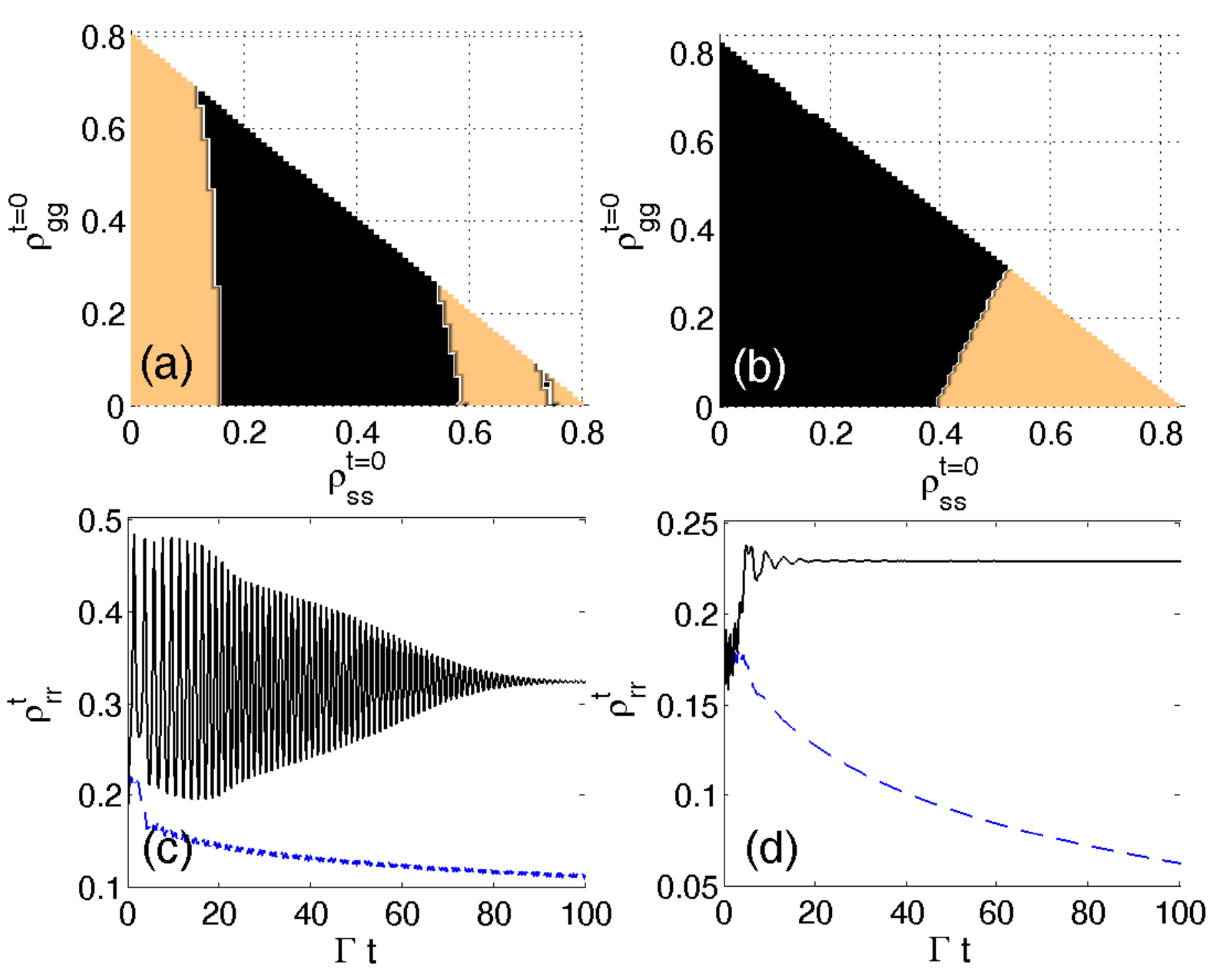}
\caption{(Color online) The
final Rydberg state population ($\protect\rho _{rr}^{t\rightarrow \infty }$) for (a) the $N$-type system and (b) the cascade system,
as a function of initial population on $s$ state ($\protect\rho _{ss}^{t=0}$) and $g$ state ($\protect\rho_{gg}^{t=0}$ ). 
Dark and light colors represent the low and high
Rydberg-state occupancy, respectively. The colorless regions are unstable and without stationary solutions.  (c) and
(d) show population evolution in the bistable region for the $N$-type system and the cascade system, respectively, where $\protect%
\rho _{gg}^{t=0}=0.1$, and $\protect\rho _{ss}^{t=0}=0.4$ (dashed blue line) and $0.7$ (solid black line). See text for other parameters.}
\label{dynamics}
\end{figure}

In the calculation, we choose $\Omega _{r}=1.0$, $\Delta =15$, other parameters are same as in Fig. \ref{Phase_diagram}(b) and (e) where the bistable steady state gives two stationary Rydberg populations
$\rho _{rr}^{t\rightarrow \infty }=0.0047$ (low) and $0.3190$ (high) for the $N$-type scheme, and $\rho _{rr}^{t\rightarrow \infty
}=0.0047$ (low) and $0.228$ (high) for the cascade scheme, respectively. 
Since the Rydberg state $\left\vert r\right\rangle $ is not subjected to the dissipation in our model, we fix its initial population by $\rho _{rr}^{t=0}=0.19$ for the $N$-type scheme and $\rho _{rr}^{t=0}=0.16$ for the cascade scheme, and then focus on the dependence of its final population on the initial population of state $\left\vert g\right\rangle $ and $\left\vert s\right\rangle $ as depicted in Fig. \ref{dynamics}(a) and (b). Note that the total particle is always conserved, $\rho _{gg}+\rho _{ee}+\rho_{ss}+\rho _{rr}=1$. 

From Fig. \ref{dynamics} we can find difference between the two schemes
in the side of small $\rho _{ss}^{t=0}$, which means state $\left\vert e\right\rangle $ and state $\left\vert
g\right\rangle $ are largely occupied at the initial time. For the cascade system whose $\gamma $ is
large and $\rho _{ss}^{t=0}$ is small, it is hard to pump atoms into the
highly excited state $\left\vert r\right\rangle $. However, in the $N$-type
system dissipations from $\left\vert e\right\rangle $ may create strong
atomic correlations between $\left\vert g\right\rangle $, $\left\vert
e\right\rangle $ and $\left\vert s\right\rangle $ (e.g. a Raman scattering
process \cite{Yuan13}), giving rise to an enhancement for population in state $\left\vert s\right\rangle $. 
It is therefore possible to realize a high Rydberg excitation. The final populations $\rho
_{jj(j=g,e,s,r)}^{t\rightarrow \infty }=(0.0974,0.0032,0.5804,0.3190)$ for the $N$%
-type scheme and $\rho _{jj(j=g,e,s,r)}^{t\rightarrow \infty
}=(0.4127,0.4589,0.1237,0.0047)$ for the cascade scheme.

On the other side when $\rho _{ss}^{t=0}$ is large, since state $\left\vert s\right\rangle $ is
directly coupled with state $\left\vert r\right\rangle $ by laser $\Omega _{r}$, the Rydberg excitation is easy. 
The final populations $\rho_{jj(j=g,e,s,r)}^{t\rightarrow \infty }=(0.0974,0.0032,0.5804,0.3190)$ for the $N$-type scheme, 
where the population in $\left\vert s\right\rangle $ and $\left\vert r\right\rangle $ are dominant. 
For the cascade scheme, due to the effect of  decay $\gamma $ and $\Gamma $ the situation is different. We have $\rho _{jj(j=g,e,s,r)}^{t\rightarrow \infty}=(0.3161,0.3506,0.1053,0.228)$, where the populations of state $\left\vert g\right\rangle $ and $\left\vert
e\right\rangle $ are not small.

Except for the above two sides, in the middle region, both of the schemes tend to a low Rydberg
occupancy with $\rho _{jj(j=g,e,s,r)}^{t\rightarrow \infty
}=(0.1373,0,0.858,0.0047)$ for the $N$-type scheme, corresponding to a dominant final occupancy of state $\left\vert s\right\rangle $, and with $\rho
_{jj(j=g,e,s,r)}^{t\rightarrow \infty }=(0.4127,0.4589,0.1237,0.0047)$
for the cascade scheme, indicating a population sharing between states $\left\vert g\right\rangle $ and $\left\vert
e\right\rangle $. 

Finally, we emphasis that even the final values of the Rydberg population are close, the dynamical evolution process and therefore the time spent on achieving the final stationary states are much different for the two schemes. In Fig. \ref{dynamics}(c) and (d), we show their different 
population dynamics for the Rydberg state in the bistable region under the same experimental
parameters. In contrast to the case of the cascade scheme, the $N$-type scheme is subjected to a strong population
oscillation while tending towards the high branch of Rydberg occupancy and
it requires more time to reach steady state. This result agrees with our former
analysis of bistability as in Fig. \ref{Phase_diagram}(b) that two non-lossy
states $\left\vert s\right\rangle $ and $\left\vert r\right\rangle $ may
lead to unstable oscillations.

\section{Experimental Realization}

Before the conclusion, we discuss the experimental realization of our proposed
schemes. The three-photon cascade scheme can not only offer an easier route for the Rydberg
excitation but also limit the excited atoms in a narrow velocity distribution, so that 
it has been adopted in recent experiments to explore the nonequilibrium phase transition in a thermal Rydberg atomic gas  \cite{Carr13}.
Especially, in this experiment the existence of intrinsic optical bistable effect has been identified as we analyzed in the above theory model. 
However, according to our theory model we find in this scheme the tri-stable phase only exists under quite rigorous parameters as implied in Fig. \ref{Phase_diagram}(e), making it a challenge to verify in experiments. As for the $N$-type scheme, it is realizable by current experimental technology and has been recently advanced in several theoretical proposals to realize Rydberg quantum simulator \cite{Weimer10} and prepare Rydberg-dressed atoms \cite{Otterbach14}.

\section{Conclusion}

In an open quantum system where driving and dissipative processes
compete with each other and settle the system on a nonequilibrium
steady state, a number of novel phenomena absent in the equilibrium system will appear \cite{Tomadin11}. 
In this work, we have investigated the effect of dissipation on the steady
state properties and dynamics for the two different excitation schemes of strong interacting Rydberg
atoms, one is of the $N$-type level structure and the other of the cascade structure.
We study the phase diagram of these schemes and find bistable and tri-stable phases in the region of strong nonlinearity caused by the Rydberg-Rydberg interaction. We investigate the population dynamics of the Rydberg state in bistable region and display its dependence of the initial population distributions. By comparing the steady state and the dynamics of these two schemes, we show their sensitive dependence on the path and the amount of the dissipation caused by the spontaneous emission \cite{Glaetzle12}. Future work will focus on the quantum state preparation \cite{Ebert14} and multi-stable switching \cite{Tony12} in the Rydberg atomic schemes by controlling and arranging the dissipation. 

J. Q. thanks P. Meystre and K. Zhang for useful discussions. This work was supported by the National Basic Research Program of China (973 Program) under Grant No. 2011CB921604, the NSFC under Grants No. 11104076 and No. 11234003, the
Specialized Research Fund for the Doctoral Program of Higher Education No.
20110076120004.


\begin{thebibliography}{99}
\bibitem{Mohapatra08} A. Mohapatra, M. Bason, B. Butscher, K. Weatherill
and C. Adams, Nat. Phys. \textbf{4} 890 (2008).

\bibitem{Peyronel12} T. Peyronel, O. Firstenberg, Q. Liang, S. Hofferberth,
A. Gorshkov, T. Pohl, M. Lukin and V. Vuleti\'{c}, Nature \textbf{488} 57 (2012).

\bibitem{WeimerS} H. Weimer Phd Dissertation University of Stuttgart (2010).

\bibitem{Saffman10} M. Saffman, T. Waller and K. M\o lmer, Rev. Mod. Phys. 
\textbf{82} 2313 (2010), and references therein.

\bibitem{Carr12} C. Carr, M. Tanasittikosol, A. Sargsyan, D. Sarkisyan, C.
Adams and K. Weatherill, Opt. Lett. \textbf{37} 3858 (2012).

\bibitem{Olmos14} B. Olmos, D. Yu and I. Lesanovsky, Phys. Rev. A \textbf{89}
023616 (2014).

\bibitem{Barreiro11} J. Barreiro, M. M\"{u}ller, P. Schindler, D. Nigg, T.
Monz, M. Chwalla, M. Hennrich, C. Roos, P. Zoller and R. Blatt, Nature 
\textbf{470} 486 (2011).

\bibitem{Genway14} S. Genway, W. Li, C. Ates, B. Lanyon and I. Lesanovsky,
Physical Review Letters \textbf{112} 023603 (2014).

\bibitem{Diehl08} S. Diehl, A. Micheli, A. Kantian, B. Kraus, H. B\"{u}chler
and P. Zoller, Nat. Phys. \textbf{4} 878 (2008).

\bibitem{Verstraete09} F. Verstraete, M. Wolf and J. Cirac, Nat. Phys. 
\textbf{5} 633 (2009).

\bibitem{Gorshkov13} A. Gorshkov, R. Nath and T. Pohl, Phys. Rev. Letts. 
\textbf{110} 153601 (2013).

\bibitem{Lemeshko13} M. Lemeshko and H. Weimer, Nat. Comm. \textbf{4} 2230
(2013).

\bibitem{Lin13} Y. Lin, J. Gaebler, F. Reiter, T. Tan, R. Bowler, A.
Sorensen, D. Leibfried and D. Wineland, Nature \textbf{504} 415 (2013).

\bibitem{Rao13} D. D. Bhaktavatsala Rao and K. K\o lmer, Phys. Rev. Letts. 
\textbf{111} 033606 (2013).

\bibitem{Carr13L} A. Carr and M. Saffman, Phys. Rev. Letts. \textbf{111}
033607 (2013).

\bibitem{Diehl10} S. Diehl, A. Tomadin, A. Micheli, R. Fazio and P. Zoller,
Phys. Rev. Letts. \textbf{105} 015702 (2010).

\bibitem{Dalla10} E. G. Dalla Torre, E. Demler, T. Giamarchi and E. Altman.
Nat. Phys. \textbf{6} 806 (2010).

\bibitem{Lesanovsky13} I. Lesanovsky and J. Garrahan, Phys. Rev. Letts. 
\textbf{111} 215305 (2013).

\bibitem{Petrosyan13} D. Petrosyan, M. H\"{o}ning and M. Fleischhauer, Phys.
Rev. A \textbf{87} 053414 (2013).

\bibitem{Lesanovsky14} I. Lesanovsky and J. Garrahan, arxiv: 1402.2126.

\bibitem{Tony11} T. Lee, H. H\"{a}ffner and M. Cross, Phy. Rev. A 
\textbf{84} 031402(R) (2011).

\bibitem{Qian12} J. Qian, G. Dong, L. Zhou and W. Zhang, Phys. Rev. A 
\textbf{85} 065401 (2012).

\bibitem{Hu13} A. Hu, T. Lee and C. Clarks, Phys. Rev. A \textbf{88}
053627 (2013).

\bibitem{Carr13} C. Carr, R. Ritter, C. Wade, C. Adams and K. Weatherill,
Phys. Rev. Letts. \textbf{111} 113901 (2013).

\bibitem{Urban09} E. Urban, T. Johnson, T. Henage, L. Isenhower, D. Yavuz,
T. Waller and M. Saffman, Nat. Phys. \textbf{5} 110 (2009).

\bibitem{Gaetan09} A. Ga\"{e}tan, Y. Miroshnychenko, T. Wilk, A. Chotia,
M. Viteau, D. Comparat, P. Pillet, A. Browaeys and P. Grangier, Nat. Phys.
\textbf{5} 115 (2009).

\bibitem{Barredo14} D. Barredo, S. Ravets, H. Labuhn, L. B\'{e}guin, A. Vernier,
F. Nogrette, T. Lahaye and A. Browaeys, Phys. Rev. Letts. \textbf{112} 183002 (2014).

\bibitem{Hakin14} A. Hankin, Y. Jau, L. Parazzoli, C. Chou, D. Armstrong, 
A. Landahl and G. Biedermann, Phys. Rev. A \textbf{89} 033416 (2014).

\bibitem{Qian13} J. Qian, L. Zhou and W. Zhang, Phys. Rev. A \textbf{87}
063421 (2013).

\bibitem{Breuer02} H. Breuer and F. Petruccione, \textit{The Theory of Open
Quantum Systems} (Oxford University Press, Oxford, 2002).

\bibitem{Comparat10} D. Comparat and P. Pillet, J. Opt. Soc. Am. B \textbf{27%
} A208 (2010).

\bibitem{Ate07} C. Ates, T. Pohl, T. Pattard and J. Rost, Phys. Rev. Letts. \textbf{98} 023002 (2007).

\bibitem{Amthor10} T. Amthor, C. Giese, C. Hofmann and M. Weidem\"{u}ller, Phys. Rev Letts. \textbf{104} 013001 (2010).

\bibitem{Ate07A} C. Ates, T. Pohl, T. Pattard and J. Rost, Phys. Rev. A \textbf{76}
013413 (2007).

\bibitem{Otterbach14} J. Otterbach and M. Lemeshko, Phys. Rev. Letts. \textbf{113} 070401 (2014).

\bibitem{Yuan13} C. Yuan, L. Chen, Z. Ou and W. Zhang, Phys. Rev. A \textbf{87}
053835 (2013).

\bibitem{Weimer10} H. Weimer, M. M\"{u}ller, I. Lesanovsky, P. Zoller and H.
B\"{u}chler, Nat. Phys. \textbf{6} 382 (2010).

\bibitem{Tomadin11} A. Tomadin, S. Diehl and P. Zoller, Phys. Rev. A \textbf{83}
013611 (2011).

\bibitem{Glaetzle12} A. Glaetzle, R. Nath, B. Zhao, G. Pupillo and P.
Zoller, Phys. Rev. A \textbf{86} 043403 (2012).

\bibitem{Ebert14} M. Ebert, A. Gill, M. Gibbons, X. Zhang, M. Saffman and T. Waller,
Phys. Rev. Letts. \textbf{112} 043602 (2014).

\bibitem{Tony12} T. Lee, H. H\"{a}ffner and M. Cross, Phys. Rev. Letts. \textbf{108}
023602 (2012).























\end{thebibliography}
\end{document}